\documentclass[12pt]{article}
\usepackage{mydraft}
\usepackage{geometry}
 \usepackage{changepage}
 \usepackage[english]{babel}
\usepackage{mathtools, wrapfig, tabularx, comment, booktabs, adjustbox}
\usepackage[blocks]{authblk}
\usepackage{rotating}

\usepackage{natbib}
\setcitestyle{authoryear,open={[},close={]}} 

\usepackage{setspace}

\title{Evaluation of Individual and Trial Level Association Metrics in the Validation of a Binary Surrogate Endpoint for a True Time-to-Event Endpoint}
\author[1]{Renee Y. Ge}
\affil[1]{Department of Biostatistics, University of North Carolina at Chapel Hill}
\author[2]{Azadeh Shohoudi}
\affil[2]{Oncology Biometrics, AstraZeneca}
\author[2]{Malini Iyengar}
\author[1]{Quefeng Li}
\author[2]{Judy Li}

\date{March 19, 2026}
\begin{document}
\maketitle

\begin{abstract}
    Candidate binary endpoints are often considered as surrogates for time-to-event (TTE) clinical endpoints, primarily because they can be assessed at earlier time points. To be submitted for regulatory approval candidate binary endpoints need to validated. The most well-known method for performing such validation employs a meta-analytic framework to estimate individual-level and trial-level association. However, the performance of these association estimates in the context of a binary surrogate has not yet been examined through a comprehensive simulation study. This research aims to systematically investigate the performance of association estimates at the trial-level and at the individual-level under various trial design choices, using both simulation studies and clinical trial data, where available.
\end{abstract}

\noindent%
\vfill

\spacingset{1.5}

\newpage
\section{Introduction}
Surrogate endpoints are endpoints that can be used as substitutes for other more direct measures of health outcomes in the evaluation of clinical interventions. In the case where a true endpoint takes too long to observe or is unethical to collect, surrogate endpoints allow for faster approval of promising new treatments. \cite{Prentice1989Surrogate} presents the first formal statistical definition of a surrogate endpoint and operational criterion to identify potential surrogates. In this work, Prentice establishes that a surrogate needs to have some predictive power of the outcome and requires both biological relevance and a statistical relationship to establish surrogacy. \cite{freedman1992} extends Prentice's work and introduces a new concept - ``proportion explained" - as the proportion of the treatment effect explained by the surrogate endpoint, and presents one of the first quantitative ways to evaluate surrogacy. While this was a step in the right direction, the proportion explained relied on strict assumptions that did not always hold in practice and was highly unstable \citep{elliot_review}. 

This led to the development of a meta-analytic approach to surrogacy evaluation by \cite{buyse_2000} wherein multiple trials are evaluated simultaneously and individual-level and trial-level associations are examined. Trial-level association is an overall measure of how well the treatment effect on the surrogate predicts the treatment effect on the true endpoint within each trial. Individual-level association is a measure of the strength of the association between the surrogate and the true endpoint, after adjusting for treatment. This meta-analytic framework was first established in the case where both the true and surrogate endpoint are continuous \citep{buyse_2000}. The initial meta-analytic framework fits a two-stage model to estimate the individual-level and trial-level associations using a fixed effects model for the surrogate and true endpoint at the first stage, and a mixed effects representation at the second stage. This work was later extended to time-to-event (TTE) surrogate and true endpoints in \cite{burz_2001} and binary surrogates for true TTE endpoints in \cite{colorectal}. In these extensions, copula models were employed to perform estimation at the first stage. A variety of other surrogacy evaluation methods have been proposed using Bayesian mixed effects models \citep{daniels1997} and within the field of causal inference \citep{gilbert2008}. Systematic simulation studies of different surrogacy evaluation methods have been performed when both the surrogate and the true endpoints are TTE \citep{shi2011, renfro2012}, but not when the surrogate is a binary endpoint.

Additionally, the FDA defines three types of surrogate endpoints based on the level of evidence available in support of the association: candidate surrogate, reasonably likely surrogate, and fully validated surrogate (FVS) \citep{FDA_Surrogate_Endpoints}. A fully validated endpoint can be accepted for full traditional approval (TA) or accelerated approval (AA), and a reasonably likely surrogate - also called an intermediate endpoint - can be accepted for AA. However, no official guidelines have been published on the amount of statistical evidence required to achieve either of these categorizations, or the recommended methods to perform such surrogacy validation. As such, researchers decide on criteria to guide their analyses that, while reasonable, are arbitrarily decided. 

In oncology clinical trials, overall survival (OS) measuring a patient's total survival time from a pre-defined time of origin is considered the gold standard clinical endpoint. However, in many hematologic malignancies in oncology, the use of OS as an endpoint presents some challenges. Increased efficacy and availability of therapies result in improvements in survival, longer required follow-up, and increased sample sizes in trials to maintain appropriate statistical power. Progression-free survival (PFS), event-free survival (EFS), and disease-free survival(DFS) are accepted surrogates for OS in hematology oncology indications, facilitating both accelerated and traditional approvals. However, as more effective salvage treatments become available and PFS times also grow longer, the need for an even earlier endpoint has become increasingly evident. 

One such endpoint that is of substantial interest as a surrogate for PFS and OS in hematology oncology is Minimal Residual Disease (MRD). MRD is a measure of the number of cancer cells remaining in a patient's peripheral blood or bone marrow after treatment. It's typically assessed as a binary endpoint of MRD undetectable/positivity if the MRD measurement is below a specified threshold (e.g. 1 cell in 10,000). Several studies have shown MRD to be significantly correlated with survival outcomes in hematological malignancies in oncology. Currently, MRD is approved for use as an intermediate endpoint for AA in B-cell acute lymphocytic leukemia (B-ALL), Philadelphia chromosome positive ALL, and multiple myeloma (MM) \citep{FDA_Surrogate_Table}. MRD was approved in MM most recently at the April 12, 2024 meeting of the FDA's Oncologic Drugs Advisory Committee (ODAC). At this meeting, the University of Miami and the International Independent Team for Endpoint Approval (i2TEAMM) separately presented their analyses on MRD as a surrogate endpoint in MM using the meta-analytic framework developed in \cite{colorectal}.

Following its approval as an intermediate endpoint in MM, interest has grown in testing MRD for other hematology-oncology indications. Given the use of the meta-analytic framework in MM approval, it is likely that industry and academic researchers will look to this analysis as a precedent in the construction of future MRD surrogacy validations in other indications. However, the impact of different trial characteristics on surrogacy metrics in the binary surrogate endpoint, and specifically MRD, setting has yet to be investigated, and it is unclear how reliably this framework can estimate the surrogacy metrics. It is often seen in practice that the magnitude of trial-level and individual-level association estimates are inconsistent, with individual-level association estimates being quite large while trial-level association metrics struggle to meet threshold values for surrogacy determination. As shown in \citep{shi2011} factors such as number of available trials, censoring rate, treatment effect size, missing data, and true association levels are likely to impact estimates, but the degree of this impact in the binary surrogate setting is unclear.

This paper aims to bridge the gap in literature by conducting a simulation-based assessment of surrogacy measures in the case of binary surrogate and true TTE endpoints. Such an investigation into the reliability of the current meta-analytic framework will offer more clarity to regulatory agencies and industry partners in the implementation of binary surrogate endpoints for true TTE endpoints and in the interpretation of surrogacy evaluation results.

The paper is organized as follows. Section 2 provides an overview of the estimation procedure for individual and trial-level association metrics. Then, the simulation settings and results are reported in Section 3. Section 4 offers an application of results in multiple myeloma (MM) clinical trial data. This is followed by a discussion of results and concluding remarks in Sections 5.

\section{Surrogacy Evaluation}
To evaluate surrogacy of a binary endpoint for a true TTE endpoint in both the simulation setting and the clinical trial data setting, the meta-analysis framework detailed by \cite{colorectal} was followed and both trial-level and individual-level associations were calculated. Individual-level association is the strength of the association between the binary surrogate and the true TTE endpoint measured through a global odds ratio (OR), Plackett copula parameter. This is defined as the odds of observing an event at a pre-specified time for the binary endpoint responders divided by the odds of observing a survival event at a pre-specified time for non-responders of the binary endpoint after adjusting for treatment effect. Trial-level association is the association between the treatment effect on the surrogate endpoint and the treatment effect on the true endpoint. The treatment effects are calculated in the form of the odds ratios of response in the treatment versus control groups on the binary surrogate endpoint and the hazard ratio of survival in the treatment versus control groups on the true TTE endpoint.

The measures of association and trial-specific treatment effects are estimated in a two-stage maximum likelihood procedure. For each of $j=1,\dots, n_i$ patients from trial $i = 1,\dots, N$ we have $(X_{ij}, \Delta_{ij}, S_{ij}, Z_{ij})$ as the failure time, censoring indictor, binary endpoint status, and treatment indicator for patient $j$ in the $i^{th}$ trial respectively. In the first stage, a logistic regression model 
$$ \text{logit}\{P(S_{ij}=0|Z_{ij})\} = \gamma_i +\alpha_iZ_{ij} $$
is used to model the binary surrogate endpoint $S_{ij}$. Here $S_{ij} = 0$ is defined as response and $S_{ij}=1$ is non-response to treatment.  A Cox proportional hazards model is used to model the observed failure time
$$\lambda_{ij}(T_{ij}|Z_{ij})=\lambda_{0i}(t_{ij})exp(\beta_iZ_{ij})$$
where $\alpha_i,\beta_i$ are the trial-specific treatment effects on the binary and TTE endpoint respectively.
Then, the Plackett copula with parameter $\theta\in \mathbb{R}_+$ is assumed to represent the joint cumulative distribution of $T_{ij}$ and $\tilde{S}_{ij}$, a latent continuous variable underlying the binary surrogate endpoint, given $Z_{ij}=z_{ij}$. This copula model describes the association between $\tilde{S}_{ij}$ and $T_{ij}$ where the parameter $\theta$ represents the aforementioned global OR (individual-level association). The likelihood function is used to obtain the estimate of $\theta$ and the trial-specific treatment effects. At the second stage, the trial-level model is used.

$$
\begin{bmatrix}
     \alpha_i \\ \beta_i
\end{bmatrix}
\sim \text{MVN} \left(
\begin{bmatrix}
   \alpha \\ \beta
\end{bmatrix},
\Sigma_{trial} = \begin{bmatrix}
     d_{aa} & d_{ab} \\
     & d_{bb}
\end{bmatrix}\right)
$$
and a version of the $R^2_{copula}$ is calculated through the dispersion matrix

\begin{equation}
R^2_{copula} = \frac{d^2_{ab}}{d_{aa}d_{bb}}
\label{eq:r2}
\end{equation}

In line with how surrogacy is typically validated in practice, this paper evaluates two additional types of $R^2$ estimators. First, a weighted least squares model is fit to the log hazard ratio of survival with the log odds ratio as the predictor. Then, either the inverse variance of the log odds ratio (corresponding to the predictor/surrogate endpoint) or the inverse sample size are used as weights, yielding $R^2_{adj}$ and $R^2_{WLS}$. 

\begin{figure}[H]
    \centering
    \includegraphics[width=1\linewidth]{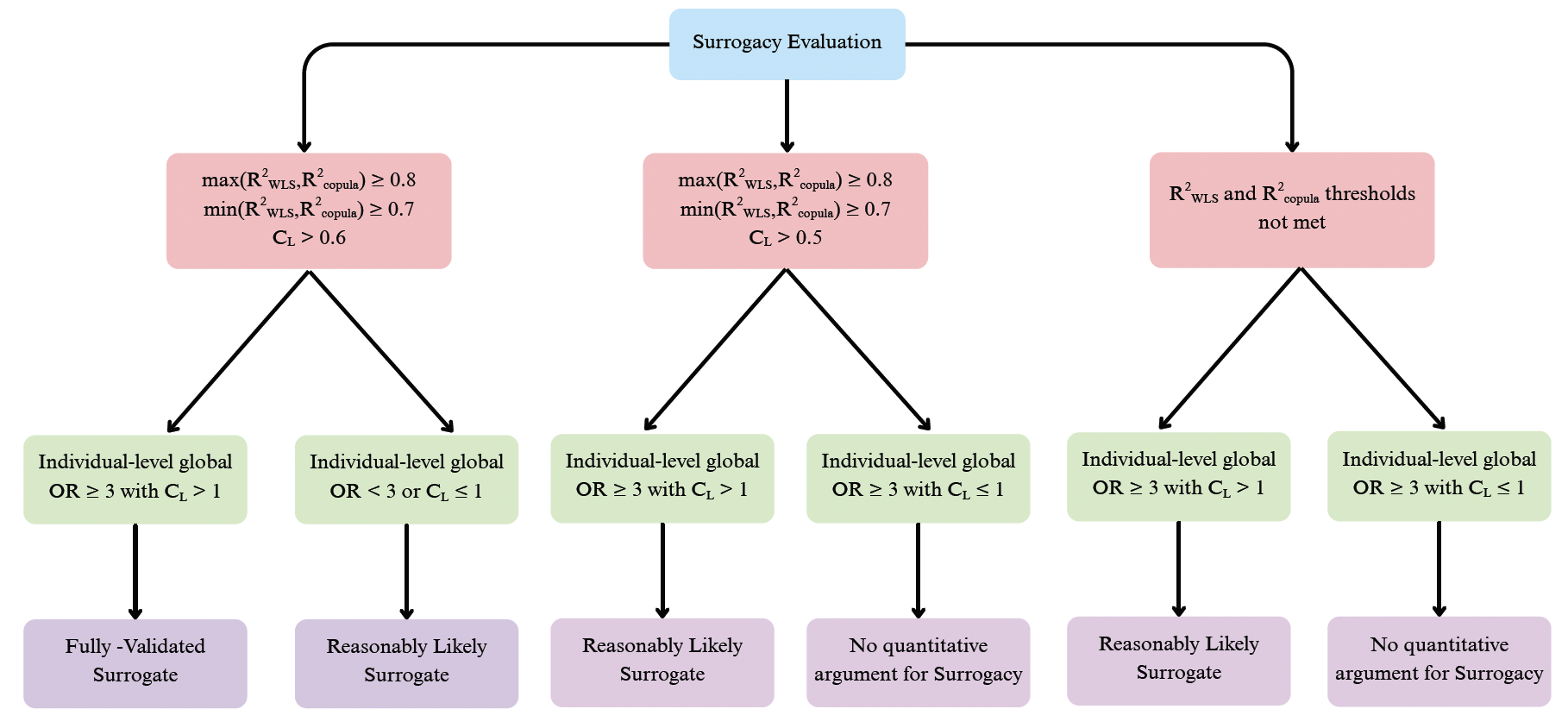}
    \caption{Surrogacy Evaluation Criterion}
    \label{fig:criteria}
\end{figure}

There is no official regulatory guidance on what values for these estimates will constitute approval of a potential surrogate endpoint as a fully validated endpoint as opposed to a reasonably likely endpoint. For the analysis that follows, the criteria implemented by the i2TEAMM in their submission of MRD as a surrogate endpoint in MM was used. This team calculated $R^2_{WLS}$, $R^2_{copula},$ and the global OR. To qualify as a FVS one of $R^2_{WLS}$ or $R^2_{copula}$ needs to be greater than 0.8 with neither being below 0.7, and with a lower 95\% confidence limit greater than 0.6. Additionally, a global OR greater than 3 with a lower 95\% confidence limit greater than 1 is required for FVS status. To qualify as a reasonably likely surrogate endpoint one of $R^2_{WLS}$ or $R^2_{copula}$ needs to be greater than 0.8 with lower 95\% confidence limit greater than 0.5 and neither estimate being below 0.7 regardless of value of the global OR. If the trial-level association does not meet this criterion, the surrogate can still qualify as a reasonably likely surrogate endpoint if the global OR is greater than 3 with a significant confidence interval. This criterion is implemented in the analysis in this paper and is depicted in Figure \ref{fig:criteria}.

\section{Simulation Study}

\subsection{Data Generation}
To reflect the two-stage evaluation process, data generation was similarly performed using a two-stage procedure to simulate data from multiple clinical trials. For each simulated trial, trial-specific treatment effects were generated from a multivariate normal distribution using a pre-specified $R^2_{copula}$. Some correlation is assumed between the baseline hazard and the odds of binary response in the control group.
    $$\begin{bmatrix}
       \gamma_i\\ \text{log}(\lambda_{0i})\\ \alpha_i\\ \beta_i
    \end{bmatrix}
    \sim \mathcal{N}(
    \begin{bmatrix}
        \gamma\\ log(\lambda_0)\\\alpha\\ \beta
    \end{bmatrix},
    \begin{bmatrix}
        1 & -\sqrt{R^2_{copula}}& 0 & 0\\ -\sqrt{R^2_{copula}} & 1& 0 & 0 \\0 & 0&1 & -\sqrt{R^2_{copula}}\\ 0 & 0 &-\sqrt{R^2_{copula}} & 1
    \end{bmatrix})$$
Then, a pair of correlated uniform random variables was generated from a Plackett copula distribution for each patient in each simulated trial based on a specified true global OR, $\theta$. Patients were assigned at a 1:1 ratio to the treatment or control groups. The first uniform random variable was transformed into the binary surrogate endpoint and the second was transformed into the true TTE endpoint under the logistic regression model and Cox proportional hazards model assumptions. The TTE endpoint was assumed to follow an exponential distribution. Transformations were performed using the inverse cumulative distribution function (CDF) for both the binary endpoint and the TTE endpoint. Assuming non-informative censoring, the censoring time was independently generated using an exponential distribution. 
All computation and data generation related to the simulations was completed using R software version 4.5.0. More details on the data generation process can be found in the appendix.
\subsection{Parameter Specification}
There are a number of factors believed to potentially impact the surrogacy metrics including the number of trials included, sample sizes within each trial, censoring rates, and strength of treatment effects. We were additionally interested in if the surrogacy metrics could be consistently estimated across a range of true association levels. Table \ref{tab:sim_factors} contains the list of factor levels considered in the simulation settings.

\begin{table}[H]
    \centering
    \small
    \caption{Simulation Factors}
    \begin{tabular}{c c} \hline
    Factors & Possible Values\\ \hline
    Trial-level $R^2_{copula}$ & 0.3, 0.65, 0.95\\
    Global OR & 1, 3, 7\\
    Number of trials & 10, 20, 30\\
    Equal sample sizes within trials & 300, 1000\\
    Unequal sample sizes & Mix of 300, 500, 1000 \\
    Censoring Rate & 5\%, 10\%, 15\% \\
    Average treatment effects $(\alpha,\beta)$ & $(0.8, -0.74), (0.3, -0.25), (1.2, -1.05)$\\\hline
  \end{tabular}
\label{tab:sim_factors}
\end{table}

We define $\alpha$ and $\beta$ as the average log OR and log HR respectively. Trial-specific baseline hazards and odds of binary response in the control group were generated from a multivariate normal distribution. There was assumed to be some correlation between the baseline hazard and odds of binary response. Parameters required for data generation were fixed at constant values derived from available real clinical trial data. Censoring times for each patient were generated from an exponential distribution by transforming the rates in the table to exponential rates.

The simulation set up was first validated to check that the data generated reflects our desired data characteristics. Then the full simulation study was conducted using the values in the table above.
For each combination of the above parameters 100 replications of the trial generation and evaluation process were performed. For each simulated set of trials, the $R^2_{copula}, R^2_{WLS}, R^2_{adj}$, and the global OR were estimated using the Surrogate package in the R programming language \citep{wim2025}. The bias, defined as $\text{estimate} -\text{truth}$, the percent change, defined as $\frac{\text{est}-\text{truth}}{\text{truth}}\times 100$, and the normalized root mean squared error (NRMSE), $\frac{\sqrt{MSE}}{\text{mean}(\text{est})}$, were recorded. The percent change and the NRMSE allow for comparisons between the global OR and the trial-level associations on the same scale. Additionally, given these estimates, the percentage of replications for each scenario where the binary endpoint would have been accepted as a fully validated endpoint or a reasonably likely surrogate endpoint were calculated. 

\subsection{Simulation Results}
We first present results averaged over all factor levels. Table \ref{tab:sim_res} presents the bias, the percent change, and the NRMSE for all four association estimates across all factor levels. When we consider variation in one factor we average over all levels of all other factors.
\begin{table}[H]

\centering
\scriptsize
\caption{Simulation results averaged over all factor levels}
\label{tab:sim_res}
\begin{tabular}{p{3cm} p{0.75cm} p{0.75cm} p{0.75cm} p{0.75cm} p{0.75cm} p{0.75cm} p{0.75cm} p{0.75cm} p{0.75cm}}

\toprule
Estimate & Bias & Perc. & NRMSE & Bias & Perc. & NRMSE & Bias & Perc. & NRMSE \\
\midrule
\multicolumn{1}{l}{\textbf{True $R^2$}} & \multicolumn{3}{c}{$R^2=0.30$} & \multicolumn{3}{c}{$R^2=0.65$} & \multicolumn{3}{c}{$R^2=0.95$} \\
\midrule
\addlinespace[0.3em]
\hspace{1em}$R^2_{copula}$ & 0.09 & 31\% & 0.57 & -0.09 & -13\% & 0.42 & -0.25 & -26\% & 0.50\\
\hspace{1em}$R^2_{WLS}$ & 0.10 & 34\% & 0.58 & -0.08 & -12\% & 0.41 & -0.24 & -25\%& 0.49\\
\hspace{1em}$R^2_{adj}$ & 0.14 & 46\% & 0.69 & -0.02 & -3\% & 0.50 & -0.19 & -19\% & 0.54\\
\hspace{1em}Global OR & 11.18 & 380\% & 0.94 & 8.57 & 303\% & 0.85 & 7.11 & 252\% & 0.80\\
\midrule
\multicolumn{1}{l}{ \textbf{True Global OR}} & \multicolumn{3}{c}{$\theta = 1$} & \multicolumn{3}{c}{$\theta = 3$} & \multicolumn{3}{c}{$\theta = 7$} \\

\midrule
\addlinespace[0.3em]
\hspace{1em}$R^2_{copula}$ & -0.10 & -5\%& 0.57 & -0.08 & -2\%& 0.49 & -0.07 & -2\% & 0.44\\
\hspace{1em}$R^2_{WLS}$ & -0.09 & -3\% & 0.56 & -0.07 & -0.1\% & 0.48 & -0.06 & 0.12\% & 0.44\\
\hspace{1em}$R^2_{adj}$ & -0.05 & 5\% & 0.67 & -0.01 & 9\% & 0.54 & 0.00 & 9\% & 0.49\\
\hspace{1em}Global OR & 4.62 & 462\% & 0.92 & 8.23 & 274\% & 0.83 & 13.99 & 200\% & 0.77\\

\midrule
\multicolumn{1}{l}{\textbf{Sample Size} } & \multicolumn{3}{c}{n = 300} & \multicolumn{3}{c}{n = 1000}& \multicolumn{3}{c}{}  \\

\midrule
\addlinespace[0.3em]

\hspace{1em}$R^2_{copula}$ & 0.00 & 10\% & 0.34 & -0.15 & -15\% & 0.67\\
\hspace{1em}$R^2_{WLS}$ & 0.00 & 12\% & 0.34 & -0.15 & -14\%& 0.66\\
\hspace{1em}$R^2_{adj}$ & 0.03 & 18\% & 0.42 & -0.09 & -4\% & 0.76\\
\hspace{1em}Global OR & 8.65 & 305\% & 0.86 & 9.39 & 322\% & 0.92\\
\midrule

\multicolumn{1}{l}{\textbf{Trial Count} } & \multicolumn{3}{c}{N = 10} & \multicolumn{3}{c}{N = 20} & \multicolumn{3}{c}{N = 30} \\

\midrule
\addlinespace[0.3em]
\hspace{1em}$R^2_{copula}$ & -0.04 & 4\% & 0.44 & -0.08 & -3\% & 0.50 & -0.12 & -9\% & 0.57\\
\hspace{1em}$R^2_{WLS}$ & -0.04 & 5\% & 0.44 & -0.07 & -1\% & 0.49 & -0.11 & -7\% & 0.55\\
\hspace{1em}$R^2_{adj}$ & 0.02 & 16\% & 0.51 & -0.03 & 7\% & 0.57 & -0.06 & 1\% & 0.64\\
\hspace{1em}Global OR & 9.14 & 319\% & 0.91 & 8.85 & 308\% & 0.88 & 8.87 & 308\% & 0.87\\
\midrule
\multicolumn{1}{l}{ \textbf{Censoring Rate}} & \multicolumn{3}{c}{Rate = 5\%} & \multicolumn{3}{c}{Rate = 10\%} & \multicolumn{3}{c}{Rate = 15\%} \\

\midrule
\addlinespace[0.3em]

\hspace{1em}$R^2_{copula}$ & -0.03 & 5\% & 0.38 & -0.08 & -3\% & 0.50 & -0.13 & -10\% & 0.63\\
\hspace{1em}$R^2_{WLS}$ & -0.03 & 6\% & 0.38 & -0.07 & -1\% & 0.49 & -0.12 & -8\% & 0.61\\
\hspace{1em}$R^2_{adj}$ & 0.01 & 12\% & 0.46 & -0.02 & 8\% & 0.57 & -0.06 & 3\% & 0.67\\
\hspace{1em}Global OR & 4.60 & 163\% & 0.63 & 8.38 & 293\% & 0.80 & 13.94 & 482\%& 0.92\\
\midrule
\multicolumn{1}{l}{\textbf{Effect Size}} & \multicolumn{3}{c}{$(\alpha,\beta) = (0.3, -0.25)$} & \multicolumn{3}{c}{$(\alpha,\beta) = (0.8, -0.74)$} & \multicolumn{3}{c}{$(\alpha,\beta) = (1.2, -1.05)$} \\

\midrule
\addlinespace[0.3em]
\hspace{1em}$R^2_{copula}$ & -0.08 & -1\% & 0.50 & -0.08 & -2\% & 0.49 & -0.09 & -5\% & 0.51\\
\hspace{1em}$R^2_{WLS}$ & -0.07 & 1\% & 0.49 & -0.07 & -0.4\% & 0.5 & -0.08 & -4\% & 0.50\\
\hspace{1em}$R^2_{adj}$ & -0.07 & -0.4\% & 0.70 & -0.01 & 11\% & 0.53 & 0.01 & 13\% & 0.48\\
\hspace{1em}Global OR & 8.73 & 304\% & 0.87 & 8.88 & 309\% & 0.89 & 9.26 & 323\% & 0.90\\
\midrule

\multicolumn{1}{l}{ \textbf{Mixed Trial Sizes}} & \multicolumn{3}{c}{N = 10} & \multicolumn{3}{c}{N = 20} & \multicolumn{3}{c}{N = 30} \\

\midrule
\addlinespace[0.3em]
\hspace{1em}$R^2_{copula}$ & -0.05 & 2\% & 0.44 & -0.09 & -5\% & 0.51 & -0.12 & -9\% & 0.59\\
\hspace{1em}$R^2_{WLS}$ & -0.04 & 4\% & 0.44 & -0.07 & -1\% & 0.49 & -0.11 & -6\% & 0.56\\
\hspace{1em}$R^2_{adj}$ & 0.03 & 16\% & 0.49 & -0.02 & 8\% & 0.55 & -0.04 & 6\%& 0.60\\
\hspace{1em}Global OR & 9.08 & 317\% & 0.90 & 8.56 & 301\% & 0.86 & 8.85 & 308\% & 0.87\\
\bottomrule
\end{tabular}
\end{table}

Generally, bias in the estimated $R^2_{copula}$ and $R^2_{WLS}$ followed similar trends. $R^2_{adj}$ consistently performed worse across bias, percent change, and the NRMSE than the other two estimates. The Global OR is overestimated throughout the table and at times was 4 times higher than the truth on average.

When varying the value of the true trial-level association, the three estimated $R^2$ metrics tended to overestimate when the true association was smaller and underestimate when the true association was greater. This is consistent with findings in \cite{shi2011} for the case where both the surrogate and true endpoint were TTE.

When the true global OR was smaller, the estimated global OR exhibited greater relative overestimation and coefficient of variation compared to when the true global OR was larger. The $R^2$ estimates did not appear to be significantly impacted by the level of the true global OR. When averaging over all factor levels, increasing the sample size per trial lead to underestimation in the trial-level association but a slight increase in the global OR estimates.

Increases in the number of trials included in the meta-analysis lead to improvements in the percent change between the estimate and the truth for the trial-level association. The bias appeared to decrease, and we observed that when the number of trials was 10 or 20 the bias pointed in a different direction than the percent change. We believed this to be a result of the underestimation
originating from trials where the true values are higher. For these values, when a percent change is calculated the actual degree of change may not be as strong, so this leads to the two averages pointing in different directions. Increasing the number of trials improved the estimation and decreased the bias, percent change, and the NRMSE for the global OR. Mixing the sample sizes within trials resulted in similar performance in each of the metrics when compared to no mixing.

For a lower censoring rate of 5\%, there was little bias in the $R^2_{copula}$ and $R^2_{WLS}$, but more bias and greater NRMSE in the $R^2_{adj}$. As the censoring rate was increased to 10\% and 15\% the $R^2$ values
tended to underestimate the truth. In contrast, the estimated global OR continued to overestimate the truth with the degree of overestimation increasing as the censoring rate increased.

Varying the average population treatment effect size did not appear to have dramatic impacts on any of the trial-level association estimates. The global OR maintained the trend of increasing overestimation as the effect size was increased.

\begin{table}[H]
\centering
\scriptsize
\caption{Surrogacy Establishment in Simulation Given True Trial-level and Individual-level Association Values}
\label{tab:surr_perc}
\begin{tabular}{cccc}
\toprule
True Trial-level Association & True Global OR & Fully Validated Surrogate & Intermediate Endpoint\\
\midrule
0.30 & 1 & 5.6\% & 98.8\%\\
0.30 & 3 & 4.2\% & 100\%\\
0.30 & 7 & 3.4\% & 100\%\\
0.65 & 1 & 23.4\% & 97.3\%\\
0.65 & 3 & 26.1\% & 100\%\\
0.65 & 7 & 24.9\% & 100\%\\
0.95 & 1 & 48.8\% & 96.5\%\\
0.95 & 3 & 69.0\% & 100\%\\
0.95 & 7 & 74.5\% & 100\%\\
\bottomrule
\end{tabular}
\end{table}
Table \ref{tab:surr_perc} presents the percentage of simulation replicates where the binary endpoint would have been established as either a FVS or an intermediate endpoint using the criterion defined in Figure \ref{fig:criteria}, stratified by the true association levels. In this study, the binary endpoint almost always would have been approved as an intermediate endpoint for the true TTE endpoint even when the true associations would not indicate such an approval. While the FVS approval was harder to obtain, this was still seen in around 5\% of simulations when the trial-level association was 0.3 and in 28\% of simulations when the trial-level association was 0.65. These are both cases where the FVS was not justified given the true values. On the other hand, in scenarios where the establishment of FVS was warranted - specifically, when the true trial-level association was 0.95 and the true global OR  was 3 or 7 - FVS status was achieved in 75\% of simulations. Within this meta-analytic framework, we observed a substantial number of both false establishments and false rejections of surrogacy. 

\begin{figure}[H]
    \centering
    \includegraphics[width=0.8\linewidth]{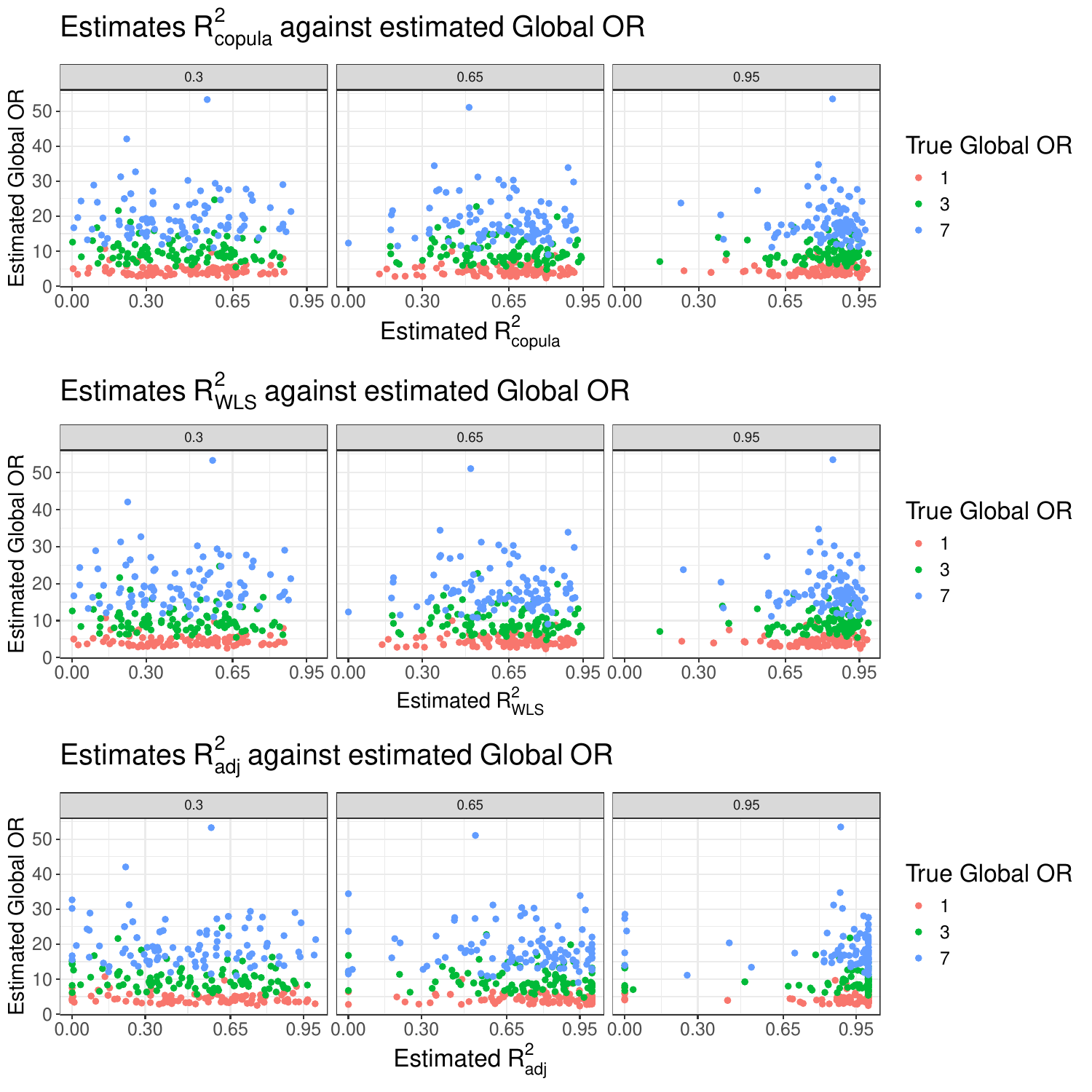}
    \caption{Scatterplots of the global OR against the three trial-level association metrics}
    \label{fig:enter-label}
\end{figure}
Next, an examination of the relationship of the estimates of global OR with those of the three $R^2$ metrics was performed. The figure above shows scatterplots of the estimated $R^2$ values on the x-axis plotted against the corresponding global OR estimate on the y-axis. In this figure, all other factors investigated in simulation were held at a constant level. 

There appeared to be greater variability in the estimated global OR when the value of the true global OR is higher. In contrast, the variability in the estimated $R^2$ values was higher when the true $R^2$
was smaller. The $R^2_{copula}$and $R^2_{WLS}$ plots were similar while the $R^2_{adj}$ plots were relatively skewed leftwards. 

The results presented thus far averaged over all factor levels; we further present two tables similar to Table 2, but specify constant factor levels. In these results, a low true level and a high true level of association were chosen. When we consider the changes across one factor, we keep all other factors at a specified constant level. 

\subsubsection{Low True Association}
We choose the constant factor levels to be $R^2 = 0.3, \text{ global }OR = 1, N = 10, \text{ sample size }=300, \text{Censoring rate = 5\%}, \text{average logOR, logHR } = (0.8, -0.74)$. Table 4 presents the results for the association estimates in this setting.
\begin{table}[H]
\centering
\scriptsize
\caption{Association estimates performance for low true association levels}

\begin{tabular}{p{3cm} p{0.75cm} p{0.75cm} p{0.75cm} p{0.75cm} p{0.75cm} p{0.75cm} p{0.75cm} p{0.75cm} p{0.75cm}}

\toprule
Estimate & Bias & Perc. & NRMSE & Bias & Perc. & NRMSE & Bias & Perc. & NRMSE \\
\midrule
\multicolumn{1}{l}{\textbf{True $R^2$}} & \multicolumn{3}{c}{$R^2=0.30$} & \multicolumn{3}{c}{$R^2=0.65$} & \multicolumn{3}{c}{$R^2=0.95$} \\
\midrule
\addlinespace[0.3em]
\hspace{1em}$R^2_{copula}$ & 0.13 & 43\% & 0.62 & -0.03 & -5\% & 0.31 & -0.25 & -27\% & 0.48\\
\hspace{1em}$R^2_{WLS}$ & 0.13 & 44\% & 0.62 & -0.03 & -5\% & 0.30 & -0.25 & -26\% & 0.47\\
\hspace{1em}$R^2_{adj}$ & 0.20 & 68\% & 0.72 & 0.08 & 12\% & 0.37 & -0.23 & -24\% & 0.65\\
\hspace{1em}Global OR & 2.91 & 291\% & 0.77 & 2.44 & 244\% & 0.72 & 2.07 & 207\% & 0.68\\
\midrule
\multicolumn{1}{l}{ \textbf{True Global OR}} & \multicolumn{3}{c}{$\theta = 1$} & \multicolumn{3}{c}{$\theta = 3$} & \multicolumn{3}{c}{$\theta = 7$} \\

\midrule
\addlinespace[0.3em]
\hspace{1em}$R^2_{copula}$ & 0.13 & 43\% & 0.62 & 0.12 & 38\% & 0.61 & 0.11 & 37\% & 0.60\\
\hspace{1em}$R^2_{WLS}$ & 0.13 & 44\% & 0.62 & 0.12 & 40\% & 0.60 & 0.11 & 38\% & 0.60\\
\hspace{1em}$R^2_{adj}$ & 0.20 & 68\% & 0.72 & 0.18 & 61\% & 0.69 & 0.17 & 56\% & 0.67\\
\hspace{1em}Global OR & 2.91 & 291\% & 0.77 & 5.28 & 176\% & 0.67 & 9.04 & 129\% & 0.61\\

\midrule
\multicolumn{1}{l}{\textbf{Sample Size} } & \multicolumn{3}{c}{n = 300} & \multicolumn{3}{c}{n = 1000}& \multicolumn{3}{c}{}  \\

\midrule
\addlinespace[0.3em]

\hspace{1em}$R^2_{copula}$ & 0.18 & 60\% & 0.59 & 0.13 & 43\% & 0.62\\
\hspace{1em}$R^2_{WLS}$ & 0.18 & 61\% & 0.60 & 0.13 & 44\% & 0.62\\
\hspace{1em}$R^2_{adj}$ & 0.20 & 68\% & 0.64 & 0.20 & 68\% & 0.72\\
\hspace{1em}Global OR & 2.91 & 291\% & 0.77 & 2.91 & 291\% & 0.77\\
\midrule

\multicolumn{1}{l}{\textbf{Trial Count} } & \multicolumn{3}{c}{N = 10} & \multicolumn{3}{c}{N = 20} & \multicolumn{3}{c}{N = 30} \\

\midrule
\addlinespace[0.3em]
\hspace{1em}$R^2_{copula}$ & 0.13 & 43\% & 0.62 & 0.11 & 37\% & 0.53 & 0.07 & 23\% & 0.46\\
\hspace{1em}$R^2_{WLS}$ & 0.13 & 44\% & 0.62 & 0.11 & 38\% & 0.53 & 0.07 & 24\% & 0.46\\
\hspace{1em}$R^2_{adj}$ & 0.20 & 68\% & 0.72 & 0.16 & 54\% & 0.65 & 0.12 & 39\% & 0.62\\
\hspace{1em}Global OR & 2.91 & 291\% & 0.77 & 2.83 & 283\% & 0.75 & 2.81 & 281\% & 0.75\\
\midrule
\multicolumn{1}{l}{ \textbf{Censoring Rate}} & \multicolumn{3}{c}{Rate = 5\%} & \multicolumn{3}{c}{Rate = 10\%} & \multicolumn{3}{c}{Rate = 15\%} \\

\midrule
\addlinespace[0.3em]

\hspace{1em}$R^2_{copula}$ & 0.13 & 43\% & 0.62 & 0.09 & 29\% & 0.66 & 0.05 & 16\% & 0.66\\
\hspace{1em}$R^2_{WLS}$ & 0.13 & 44\% & 0.62 & 0.10 & 32\% & 0.65 & 0.06 & 20\% & 0.65\\
\hspace{1em}$R^2_{adj}$ & 0.20 & 68\% & 0.72 & 0.18 & 60\% & 0.79 & 0.11 & 34\%& 0.92\\
\hspace{1em}Global OR & 2.91 & 291\% & 0.77 & 5.19 & 519\% & 0.87 & 9.53 & 953\% & 0.96\\
\midrule
\multicolumn{1}{l}{\textbf{Effect Size} } & \multicolumn{3}{c}{$(\alpha,\beta) = (0.3, -0.25)$} & \multicolumn{3}{c}{$(\alpha,\beta) = (0.8, -0.74)$} & \multicolumn{3}{c}{$(\alpha,\beta) = (1.2, -1.05)$} \\

\midrule
\addlinespace[0.3em]
\hspace{1em}$R^2_{copula}$ & 0.16 & 54\% & 0.59 & 0.13 & 43\% & 0.62 & 0.12 & 39\% & 0.63\\
\hspace{1em}$R^2_{WLS}$ & 0.17 & 56\% & 0.59 & 0.13 & 44\% & 0.62 & 0.12 & 40\% & 0.63\\
\hspace{1em}$R^2_{adj}$ & 0.18 & 60\% & 0.78 & 0.20 & 68\% & 0.72 & 0.22 & 73\% & 0.73\\
\hspace{1em}Global OR & 2.78 & 278\% & 0.76 & 2.91 & 291\% & 0.77 & 3.10 & 310\% & 0.78\\
\midrule

\multicolumn{1}{l}{ \textbf{Mixed Trial Sizes}} & \multicolumn{3}{c}{N = 10} & \multicolumn{3}{c}{N = 20} & \multicolumn{3}{c}{N = 30} \\

\midrule
\addlinespace[0.3em]
\hspace{1em}$R^2_{copula}$ & 0.12 & 41\% & 0.65 & 0.12 & 39\% & 0.50 & 0.11 & 37\% & 0.51\\
\hspace{1em}$R^2_{WLS}$ & 0.13 & 43\% & 0.66 & 0.13 & 42\% & 0.53 & 0.11 & 38\% & 0.54\\
\hspace{1em}$R^2_{adj}$ & 0.17 & 57\% & 0.71 & 0.15 & 51\% & 0.62 & 0.20 & 66\% & 0.60\\
\hspace{1em}Global OR & 2.95 & 295\% & 0.78 & 2.71 & 271\% & 0.74 & 2.89 & 289\% & 0.75\\
\bottomrule
\end{tabular}
\end{table}

Here we can see much more specific trends in the estimates. When looking at the different levels of the true $R^2$ we again see that the estimation was most accurate for a true association level set at 0.65 (moderate but still high). The NRMSE was additionally the smallest in this case indicating that estimates were more concentrated around the mean than for the other two true $R^2 $ levels. Estimation in the global OR also appeared to improve as the true trial level association increases. The global OR performed better in the bias for lower true levels, and estimation of the $R^2 $ metrics improved with increasing global OR. 

When the trial sample size was increased, reductions in the bias in the $R^2_{copula}$ and $R^2_{WLS}$ estimates were observed. Global OR appeared unaffected by sample size when the true association was low. There appeared to be more variation relative to the mean when sample size was increased. When the number of trials included in each meta-analysis was increased, there were clear improvements in the bias, percent change, and NRMSE for all metrics with $R^2_{copula}$ being the best performer of the three $R^2$ estimates.

Here, we can also more clearly elucidate the impact of the treatment effect size where stronger assumed population average logOR and logHR appeared to decrease the bias in the $R^2$ estimates, but increased bias in the global OR. Unequal trial sample sizes had no real impact on estimation performance when compared to keeping sample sizes per trial equal. 

\subsubsection{High True Association}
\begin{table}[H]
\centering
\scriptsize
\caption{Association estimates performance for high true association levels}

\begin{tabular}{p{3cm} p{0.75cm} p{0.75cm} p{0.75cm} p{0.75cm} p{0.75cm} p{0.75cm} p{0.75cm} p{0.75cm} p{0.75cm}}

\toprule
Estimate & Bias & Perc. & NRMSE & Bias & Perc. & NRMSE & Bias & Perc. & NRMSE \\
\midrule
\multicolumn{1}{l}{\textbf{True $R^2$}} & \multicolumn{3}{c}{$R^2=0.30$} & \multicolumn{3}{c}{$R^2=0.65$} & \multicolumn{3}{c}{$R^2=0.95$} \\
\midrule
\addlinespace[0.3em]
\hspace{1em}$R^2_{copula}$ & 0.11 & 37\%  & 0.60 & -0.03 & -4\%  & 0.29 & -0.19 & -20\%  & 0.35\\
\hspace{1em}$R^2_{WLS}$ & 0.11 & 38\%  & 0.60 & -0.02 & -4\%  & 0.29 & -0.19 & -20\%  & 0.35\\
\hspace{1em}$R^2_{adj}$ & 0.17 & 56\%  & 0.67 & 0.07 & 11\%  & 0.35 & -0.12 & -12\%  & 0.40\\
\hspace{1em}Global OR & 9.04 & 129\%  & 0.61 & 6.91 & 99\%  & 0.52 & 5.73 & 82\%  & 0.47\\
\midrule
\multicolumn{1}{l}{ \textbf{True Global OR}} & \multicolumn{3}{c}{$\theta = 1$} & \multicolumn{3}{c}{$\theta = 3$} & \multicolumn{3}{c}{$\theta = 7$} \\

\midrule
\addlinespace[0.3em]
\hspace{1em}$R^2_{copula}$ & -0.25 & -27\%  & 0.48 & -0.19 & -20\%  & 0.33 & -0.19 & -20\% & 0.35\\
\hspace{1em}$R^2_{WLS}$ & -0.25 & -26\%  & 0.47 & -0.19 & -20\%  & 0.32 & -0.19 & -20\%  & 0.35\\
\hspace{1em}$R^2_{adj}$ & -0.23 & -24\%  & 0.65 & -0.10 & -10\% & 0.36 & -0.12 & -12\% & 0.40\\
\hspace{1em}Global OR & 2.07 & 207\%  & 0.68 & 3.54 & 118\%  & 0.56 & 5.73 & 82\%  & 0.47\\

\midrule
\multicolumn{1}{l}{\textbf{Sample Size} } & \multicolumn{3}{c}{n = 300} & \multicolumn{3}{c}{n = 1000}& \multicolumn{3}{c}{}  \\

\midrule
\addlinespace[0.3em]

\hspace{1em}$R^2_{copula}$ & -0.05 & -5\% & 0.10 & -0.19 & -20\%  & 0.35\\
\hspace{1em}$R^2_{WLS}$ & -0.05 & -5\%  & 0.10 & -0.19 & -20\%  & 0.35\\
\hspace{1em}$R^2_{adj}$ & -0.01 & -2\%  & 0.15 & -0.12 & -12\%  & 0.40\\
\hspace{1em}Global OR & 5.06 & 72\%  & 0.43 & 5.73 & 82\% & 0.47\\
\midrule

\multicolumn{1}{l}{\textbf{Trial Count} } & \multicolumn{3}{c}{N = 10} & \multicolumn{3}{c}{N = 20} & \multicolumn{3}{c}{N = 30} \\

\midrule
\addlinespace[0.3em]
\hspace{1em}$R^2_{copula}$ & -0.19 & -20\% & 0.35 & -0.21 & -23\% & 0.40 & -0.26 & -28\%  & 0.51\\
\hspace{1em}$R^2_{WLS}$ & -0.19 & -20\% & 0.35 & -0.21 & -22\% & 0.40 & -0.26 & -27\% & 0.51\\
\hspace{1em}$R^2_{adj}$ & -0.12 & -12\% & 0.40 & -0.14 & -14\% & 0.42 & -0.18 & -19\% & 0.50\\
\hspace{1em}Global OR & 5.73 & 82\% & 0.47 & 5.41 & 77\% & 0.45 & 5.40 & 77\% & 0.44\\
\midrule
\multicolumn{1}{l}{ \textbf{Censoring Rate}} & \multicolumn{3}{c}{Rate = 5\%} & \multicolumn{3}{c}{Rate = 10\%} & \multicolumn{3}{c}{Rate = 15\%} \\

\midrule
\addlinespace[0.3em]

\hspace{1em}$R^2_{copula}$ & -0.19 & -20\% & 0.35 & -0.27 & -29\% & 0.54 & -0.30 & -32\% & 0.58\\
\hspace{1em}$R^2_{WLS}$ & -0.19 & -20\% & 0.35 & -0.27 & -28\% & 0.53 & -0.30 & -31\% & 0.56\\
\hspace{1em}$R^2_{adj}$ & -0.12 & -12\% & 0.40 & -0.18 & -18\% & 0.52 & -0.19 & -20\% & 0.56\\
\hspace{1em}Global OR & 5.73 & 82\% & 0.47 & 10.90 & 156\% & 0.63 & 20.97 & 300\% & 0.83\\
\midrule
\multicolumn{1}{l}{\textbf{Effect Size} } & \multicolumn{3}{c}{Small} & \multicolumn{3}{c}{Medium} & \multicolumn{3}{c}{High} \\

\midrule
\addlinespace[0.3em]
\hspace{1em}$R^2_{copula}$ & -0.17 & -18\% & 0.33 & -0.19 & -20\% & 0.35 & -0.18 & -19\%& 0.32\\
\hspace{1em}$R^2_{WLS}$ & -0.17 & -18\% & 0.32 & -0.19 & -20\% & 0.35 & -0.18 & -19\% & 0.32\\
\hspace{1em}$R^2_{adj}$ & -0.13 & -14\% & 0.44 & -0.12 & -12\%& 0.40 & -0.08 & -9\% & 0.30\\
\hspace{1em}Global OR & 5.40 & 77\% & 0.45 & 5.73 & 82\% & 0.47 & 6.09 & 87\% & 0.49\\
\midrule

\multicolumn{1}{l}{ \textbf{Mixed Trial Sizes}} & \multicolumn{3}{c}{N = 10} & \multicolumn{3}{c}{N = 20} & \multicolumn{3}{c}{N = 30} \\

\midrule
\addlinespace[0.3em]
\hspace{1em}$R^2_{copula}$ & -0.13 & -14\% & 0.25 & -0.14 & -14\% & 0.26 & -0.17 & -17\% & 0.32\\
\hspace{1em}$R^2_{WLS}$ & -0.12 & -13\% & 0.24 & -0.12 & -13\% & 0.25 & -0.15 & -16\% & 0.30\\
\hspace{1em}$R^2_{adj}$ & -0.07 & -7\% & 0.29 & -0.05 & -6\% & 0.27 & -0.10 & -10\% & 0.35\\
\hspace{1em}Global OR & 5.44 & 78\% & 0.45 & 5.10 & 73\% & 0.43 & 5.16 & 74\% & 0.43\\
\bottomrule
\end{tabular}
\end{table}
In this section we discuss results when the true association level was high, more explicitly when the true trial-level association was set to 0.95 and the true global OR was set to 9.

Here, the $R^2$ was underestimated in general. Increasing the trial sample size from 300 to 1000 increased the degree of underestimation from around $5\%$ to $20\%$. Increasing the trial count and the censoring rates had a similar impact of increasing the degree of underestimation in the $R^2$. The global OR estimation improved with increased trial count and for smaller censoring rates. Global OR seemed to be most affected by censoring rate out of all factors considered.

\section{Clinical Trial Data Application}
In this section, we present an application of these simulation results in MM clinical trials. We conducted a literature search of the PubMed database for Phase 2 and 3 clinical trials in MM. We additionally reviewed the trials included in the analysis of MRD as a surrogate for survival at the FDA ODAC meeting. Eleven trials were selected based on this review with selection being based on the availability of Kaplan-Meier (KM) plots reporting survival curves stratified by MRD and treatment status in the publications associated with each trial. The trials selected were CLARION \citep{NCT01818752}, TOURMALINE-MM3 \citep{NCT02181413}, ALCYONE \citep{NCT02195479}, MAIA \citep{NCT02252172}, CASSIOPEIA \citep{NCT02541383}, GRIFFIN \citep{NCT02874742}, OCTANS \citep{NCT03217812}, IKEMA \citep{NCT03275285}, IMROZ \citep{NCT03319667}, GMMG-HD7 \citep{NCT03617731}, and AURIGA \citep{NCT03901963}. Details on the trials are presented in Table \ref{tab:trials}.

\begin{sidewaystable}[]
    \centering
    \caption{Selected Multiple Myeloma Clinical Trials }

    \scriptsize
    \begin{tabular}{|c|p{2cm}|p{2cm}|p{1cm}|p{1.5cm}|p{1cm}|c|p{3cm}|c|}
    \toprule
    &NCT & Trial Name & Study Type & Patient Population & Sample Size & MRD Assessment & MRD Timing & MRD Sensitivity   \\
    \midrule
     1 & NCT01818752 & CLARION & Phase 3 & Transplant-ineligible NDMM & 223 & Flow cytometry & End of treatment & $2\times10^{-6}$\\
     2 & NCT02181413 & TOURMALINE-MM3 & Phase 3 & & 556 & Flow cytometry& 13 months & $1\times10^{-5}$\\

     3 & NCT02195479 & ALCYONE & Phase 3 & Transplant-ineligible NDMM & 706 & NGS & 12, 18, 24, 30 months after first treatment dose & $1\times10^{-5}$\\
     4 & NCT02252172 & MAIA & Phase 3 & Transplant-ineligible NDMM & 737 & NGS & 12, 18, 24, 30, 36, 48, and 60 months after the first treatment dose & $1\times10^{-5}$\\
     5 & NCT02541383 & CASSIOPEIA & Phase 3 & Transplant-eligible NDMM & 1040 & NGS & End of induction & $1\times10^{-5}$\\
     6 & NCT02874742 & GRIFFIN & Phase 2 & Transplant-eligible NDMM & 220 & NGS & End of consolidation & $1\times10^{-5}$\\
     7 & NCT03217812 & OCTANS & Phase 3 & Transplant ineligible NDMM & 207 & Flow cytometry & 12, 18, 24, and 30 months after the first treatment dose & $1\times10^{-5}$\\
     8 & NCT03275285 & IKEMA & Phase 3 & Relapsed/ refractory MM & 302 & NGS & After complete response or very good partial response & $1\times10^{-5}$\\
     9 & NCT03319667 & IMROZ & Phase 3 & Transplant-ineligible NDMM & 446 & NGS & End of induction & $1\times10^{-5}$\\
     10 & NCT03617731 & GMMG-HD7 & Phase 3 & Transplant-eligible NDMM & 586 & Flow cytometry & 4-5 months after randomization & $1\times10^{-5}$\\
     11 & NCT03901963 & AURIGA & Phase 3 & Post-transplant MM & 200 & NGS & 12 months after randomization &$1\times10^{-5}$\\
     \bottomrule
    \multicolumn{7}{l}{NDMM: Newly diagnosed multiple myeloma, NGS: Next-gen sequencing, Flow: flow cytometry}
    \end{tabular}
    \label{tab:trials}
\end{sidewaystable}

\begin{figure}
    \centering
    \caption{Forest plot of HR in MM clinical trials}
    \includegraphics[width=0.8\linewidth]{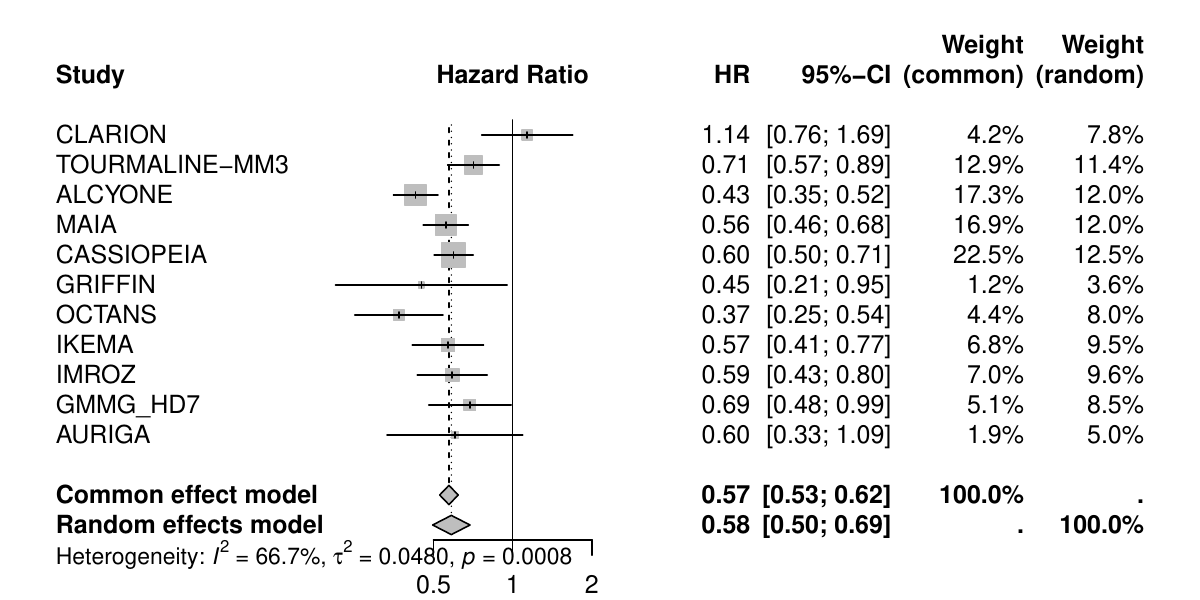}
    
    \label{fig:forest}
\end{figure}

\begin{figure}
    \centering
    \caption{Forest plot of OR in MM clinical trials}
    \includegraphics[width=0.8\linewidth]{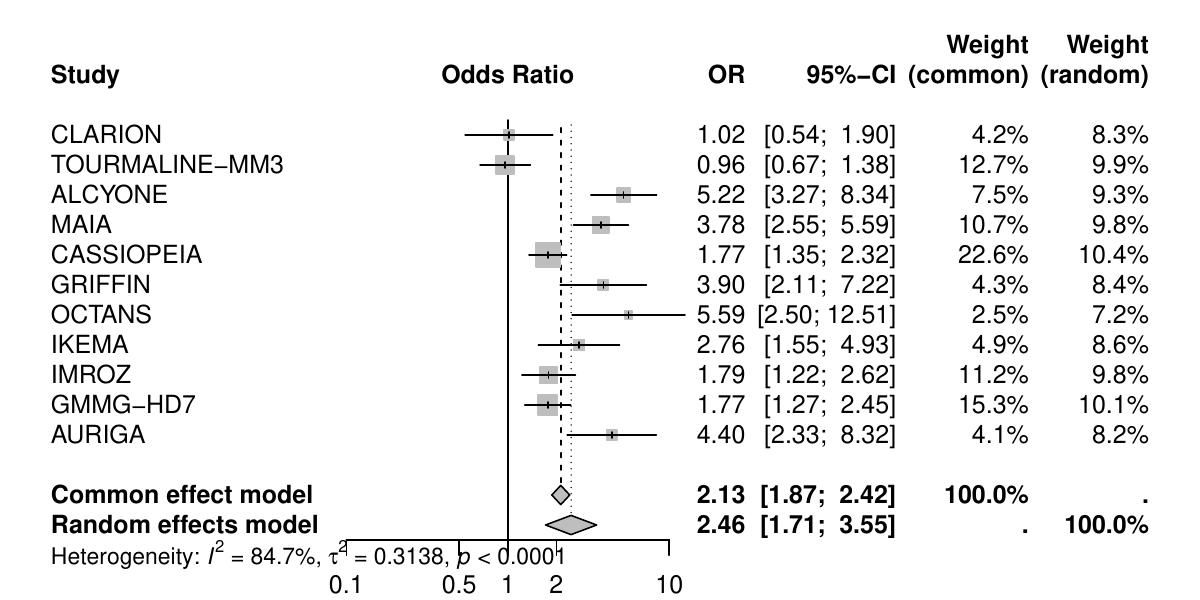}
    
    \label{fig:forest_OR}
\end{figure}

\begin{figure}
    \centering
    \caption{Forest plot of HR for MRD in MM clinical trials}
    \includegraphics[width=0.8\linewidth]{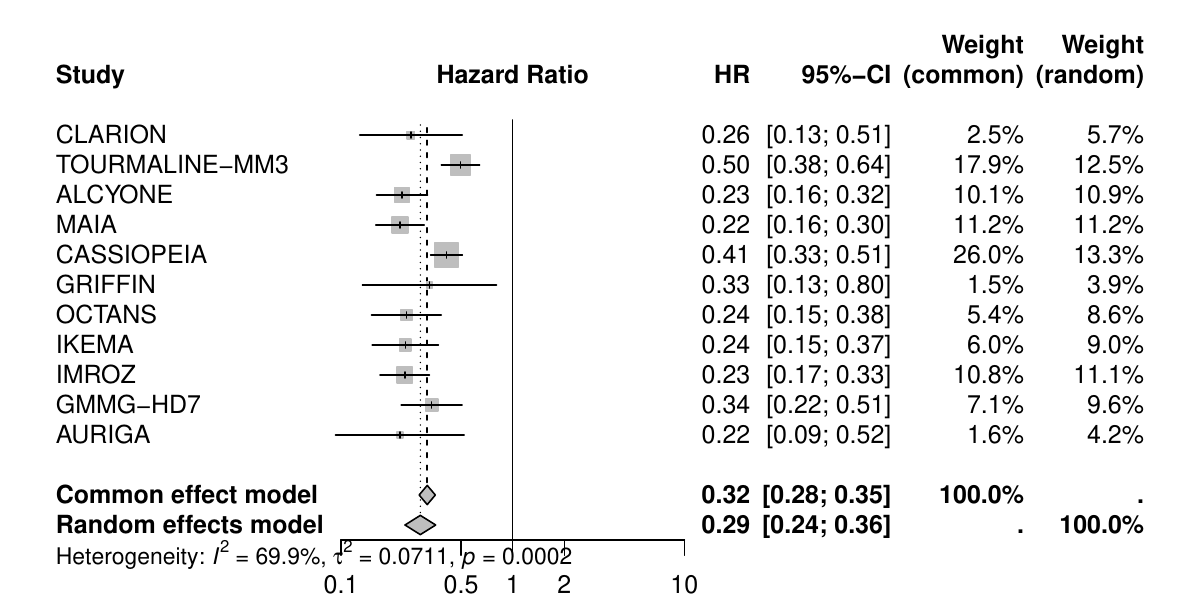}
    
    \label{fig:forest_HR_mrd}
\end{figure}

\begin{figure}
    \centering
    \caption{Treatment Effects on PFS vs. MRD in selected MM clinical trials}
    \includegraphics[width=0.7\linewidth]{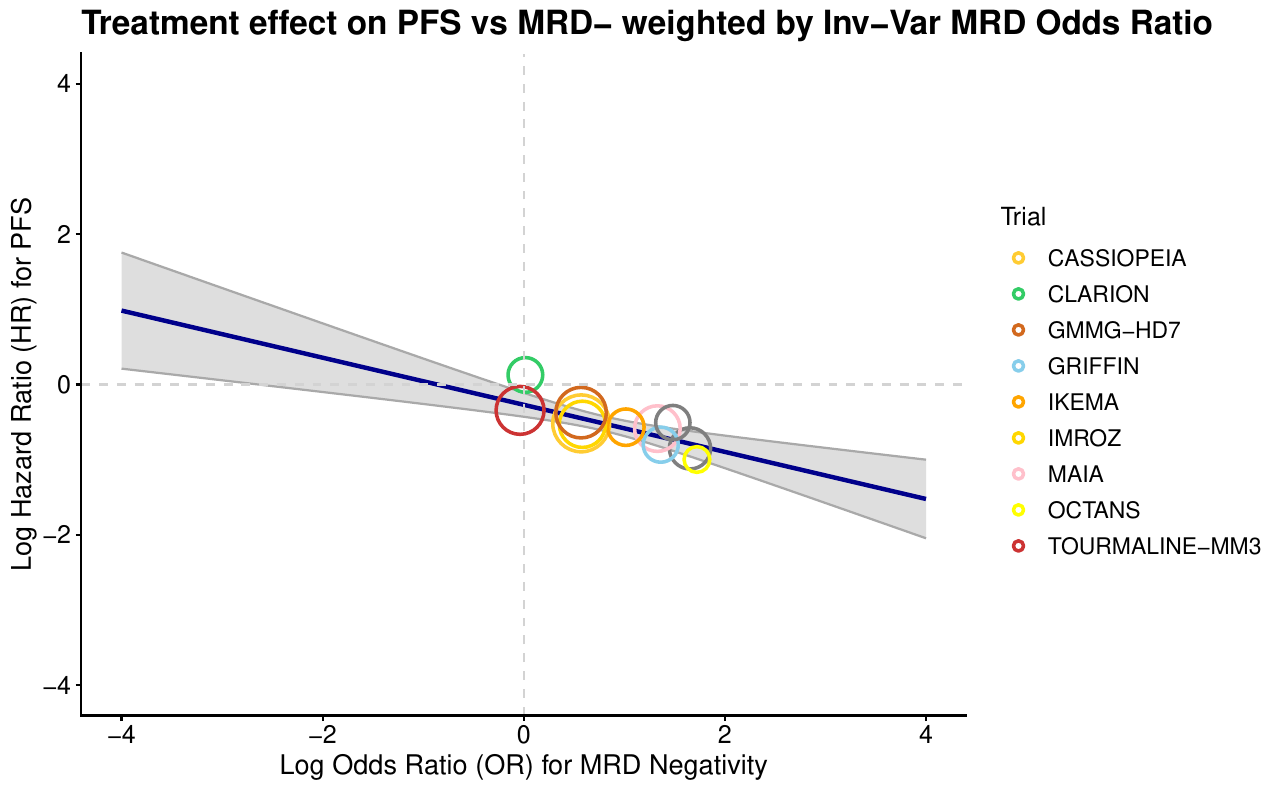}
    
    \label{fig:te}
\end{figure}

For each trial, survival curves were extracted from associated publications and data points from the curves were recorded using the ``ScanIt" free software which uses KM plot images as input and outputs coordinates for the curves. These coordinates were then converted to individual patient data (IPD) using the R package ``IPDfromKM" \citep{liu2021}. To validate the generated data, the survival curves were reconstructed and compared to the original plots - visually validated (not part of this article). Figures \ref{fig:forest}-\ref{fig:forest_HR_mrd} report estimated HRs of PFS for treatment, estimated ORs of MRD negativity for treatment, and estimated HRs of PFS by MRD Status and the corresponding 95\% confidence intervals, in forest plots. The majority of trials had significant HRs for treatment below 1 except for CLARION which had an estimated HR of 1.14. Similarly, most trials had significant ORs for treatment except CLARION and TOURMALINE-MM3. Figure \ref{fig:te} plots the treatment effects on PFS against the treatment effects on MRD, with the fitted regression model. The plotted points were clumped together indicating that the trials selected exhibited relatively similar HRs and ORs. 

In this analysis, we considered MRD and PFS to be the surrogate and true endpoints respectively. Sample sizes within each trial varied from 200 - 1040. The average logOR and logHR in the trials was 0.93 and -0.54 respectively. The censoring rate was generally around 3\%. 

A separate set of simulations was performed using parameters specific to the data in these eleven trials. We set trial sizes, censoring rate, and treatment effect sizes to reflect the range observed in these trials. We set the true trial-level association to be 0.7 and the true global OR to be 4 in accordance with the estimated result from the included trials. All other data generation parameters were fixed at estimates from the real trial data. 500 simulations were performed using the above fixed parameters. The results of these simulations are given in Table \ref{tab:real_data}.

\begin{table}[H]

\centering
\scriptsize
\caption{Real Data Estimation and Simulation Result Comparison}
\label{tab:real_data}

\begin{tabular}{p{3cm} p{0.9cm} p{1.5cm} p{0.75cm} p{0.75cm} p{0.75cm} p{0.75cm}}
\toprule
\multicolumn{1}{l}{} & \multicolumn{2}{c}{Real data estimation} & \multicolumn{2}{c}{Simulation results}  \\
\midrule
Surrogacy Metrics & Estimate & 95\% CI & Average & NRMSE  \\
\midrule
\addlinespace[0.3em]
\hspace{1em}$R^2_{copula}$ & 0.68 & (0.28, 0.89)  & 0.77 & 0.188 \\
\hspace{1em}$R^2_{WLS}$ & 0.69 & (0.28, 0.89)  & 0.78 & 0.19 \\
\hspace{1em}$R^2_{adj}$ & 0.998 & (0.33, 1)  & 0.8 & 0.254  \\
\hspace{1em}Global OR & 4.37 & (3.78, 4.96)  & 13 & 0.947  \\

\bottomrule
\end{tabular}
\end{table}

The simulation results showed a slight overestimation of the true trial-level association, which was set at 0.7. The adjusted $R^2$ estimate in the real data was much higher than the $R^2_{WLS}$ and $R^2_{copula}$ estimates. $R^2_{adj}$ also performed the worst out of the trial-level association estimates in the simulation setting. Given that the trial-level association estimates slightly overestimated the truth in simulation, the estimated trial-level association we see in the real data may also overestimate the true underlying association, but remains fairly accurate. The global OR is also severely overestimated in the simulation setting. Similarly, this indicates that the true individual-level association may be smaller in magnitude than was estimated in the real data. 

We did not see great variability in the logHR and logOR estimates in the real trials. Previous results have indicated that a wider range of HR and OR among the included trials could improve surrogacy estimation. 

\section{Discussion}
While large-scale simulation studies have been performed in the setting where both the surrogate and true endpoints were TTE, this is the first time such an investigation has been performed in the case of binary surrogate endpoints for true TTE endpoints. When conducting surrogacy validation analyses in practice, the analysis is influenced by a number of factors including the number of available trials, variability in the treatment effect, censoring rate, and varying sample sizes across trials. Examining these factors in simulation, along with an investigation in how different true levels of association may impact estimates, is important to guide the interpretation of such evaluation results. The data generation procedure developed, which parallels the analysis framework, allowed us great control over the simulation set up and the ability to test these factors. 

This result demonstrates that the global OR is consistently overestimated across all scenarios evaluated. Trial-level associations can be estimated with greater accuracy when censoring rates are low. Furthermore, increasing both the number of trials and sample sizes enhances the performance of each association metric. These conclusions are reflected in our application to MM clinical trial data. However, it is noted that our results are not expected to be a reproduction of those presented at the FDA ODAC Meeting on MRD in MM as our trial inclusion criteria is dependent on KM plot availability.

In our simulation setting, censoring was generated independently of the survival time obtained from the copula model. We treated the survival time obtained from the copula model as the true failure time and obtained the observed time by taking the minimum of this failure time and the censoring time. However, the meta-analytic framework models the observed time and not the true unknown failure time for each participant. We don't believe this choice in our simulation would strongly impact the estimation results as independent censoring is a common and reasonable assumption in simulation studies. 

For all scenarios in the simulation study, the time of binary endpoint assessment was chosen to be at six months. This choice is a common assessment timepoint for early binary endpoints in hematology oncology indications. However, binary surrogate endpoints raise the issue of length bias, whereby patients that have longer survival time are also more likely to be responders (in terms of the binary surrogate endpoint). This is one explanation for the degree of bias observed in the global OR in these simulations. The impact of binary endpoint assessment timing and the need for landmark analyses has not been systematically studied in the surrogacy evaluation framework and is an interesting avenue for possible future work. Given this knowledge, we may assume some degree of bias can be attributed to the delayed binary response timing, although the magnitude of such bias is unknown. This methodology is currently being investigated and recommendations for landmark analyses in the surrogacy validation setting will be published in a separate manuscript.

Binary endpoint assessment timing aside, there are a number of conclusions that can be drawn from this study. First, $R^2_{copula}$ and $R^2_{WLS}$ appear to be more consistent measures of trial-level association than $R^2_{adj}$. This was observed to be the case in almost all factor combinations. Additionally, the global OR appears to severely overestimate the true level of association, even when no association exists. This opens the door to future work and new methodology which can better handle bias in estimation. However, it is still undeniable that surrogate endpoints are a pivotal tool that allow patients to more quickly receive novel life-saving treatment. While these new methodologies are being developed, readjusting the thresholds for the current individual-level and trial-level association metrics would serve as an appropriate balance of patient benefit and caution of statistical bias.


\section*{Acknowledgments}
The authors would like to thank Oncology Biometrics at AstraZeneca PLC for the research funding.

\section*{Conflicts of Interest}
None to report.

\section*{Author Contributions}
R.Y.G., A.S., M.I., and J.L., contributed to the design and implementation of experiments. R.Y.G., A.S., M.I., Q.L., and J.L. contributed to the writing of the manuscript.

\section*{Appendix A: Generating Data from the bivariate Plackett copula}
For a bivariate random variable $Y=(Y_1,Y_2)$ with joint distribution function $F(y_1, y_2)$ and marginal distributions $F_1(y_1), F_2(y_2)$. The joint distribution function can be written in the form of a copula function 

$$F(y_1, y_2) = C_\theta\{F_1(y_1), F_2(y_2),\theta\}$$
where $\theta$ is the parameter of the Plackett copula. The Plackett copula function is defined as

$$
    C_\theta(u_1,u_2,\theta) =
    \begin{cases}
      \frac{1+(u_1+u_2)(\theta-1-S_\theta(u_1,u_2))}{2(\theta-1)} & \text{if } \theta \neq 1\\
      0 & \text{otherwise}
    \end{cases} 
$$
and
$$
S_\theta(u_1,u_2)=\sqrt{\{1+(\theta-1)(u_1+u_2)\}^2+4\theta(1-\theta)u_1u_2}
$$
If we can generate a sample $(U_1,U_2)$ from the copula function  then the required surrogate and true endpoint pair can be constructed as $(S,T)=(F_S^{-1}(U_1), F_T^{-1}(U_2))$. Generating variables from the copula can be done by iterative conditioning \citep{bouye2007, yan2007}
\begin{enumerate}
    \item Generate independent uniforms $V_1\sim Uniform(0,1)$ and $V_2\sim Uniform(0,1)$.
    \item Then we use the conditional cdf of the copula to simulate $U_2$.
    $$ U_1=V_1$$
    $$\partial_{u_1}C_{(u_1,u_2)}(u_2)= P(U_2\leq u_2| U_1 = u_1) =v_2$$
    \item We invert the above and solve:
    $$ u_2 = \partial_{u_1}^{-1}C_{(u_1,u_2)}(v_2)$$
    \item Return the pair $(U_1, U_2)$ and use the classical inverse cdf transformation to obtain the surrogate and true endpoint pair.
\end{enumerate}

\section*{Appendix B: Generating Data for Simulation}
\begin{enumerate}
    \item Generate $(\gamma_i, \text{log}(\lambda_{0i}), \alpha_i,\beta_i)$ from a multivariate normal distribution 
    
    $$\begin{bmatrix}
       \gamma_i\\ \text{log}(\lambda_{0i})\\ \alpha_i\\ \beta_i
    \end{bmatrix}
    \sim \mathcal{N}(
    \begin{bmatrix}
        \gamma\\ log(\lambda_0)\\\alpha\\ \beta
    \end{bmatrix},
    \begin{bmatrix}
        1 & -\sqrt{R^2_{trial}}& 0 & 0\\ -\sqrt{R^2_{trial}} & 1& 0 & 0 \\0 & 0&1 & -\sqrt{R^2_{trial}}\\ 0 & 0 &-\sqrt{R^2_{trial}} & 1
    \end{bmatrix})$$
    \item For each trial $i$, generate $n_i$ pairs of uniform random variables $(u_{ij1}, u_{ij2})$ from a Plackett Copula distribution with parameter $\Theta$
    \item Assign participants with 1:1 ratio to treatment:control and define $Z_{ij}$ as an indicator for whether participant $j$ in the $i^{th}$ trial is in the treatment group.
        \item For each trial $i$, define $b_0 = \text{exp}(\gamma_i)$ and $p_0=\frac{b_0}{1+b_0}$ as the odds of MRD negativity in the control group
    \item Define $b_1 = b_0exp(\alpha_i)$ and $p_1=\frac{b_1}{1+b_1}$ as the odds and the probability of MRD negativity respectively in the treatment group
    \item Let $S_{ij}$ be the MRD Status for the $j^{th}$ participant in the $i^{th}$ trial
    \begin{enumerate}
        \item If $u_{ij1}<p_0^{1-z_{ij}}p_1^{z_{ij}}$, set $S_{ij}=0$
        \item Calculate survival time by $T_{ij} =-\frac{log(u_2)}{\lambda_{0i}(exp(\beta_i))^{z_{ij}}}$
    \end{enumerate}
    \item Let $\lambda_c=-log(1-r_c)$. Generate true censoring time, $C_{ij}$ for all participants from $Exp(\lambda_c)$ distribution
    \item Let $X_{ij}=T_{ij}\wedge C_{ij}$ and $\Delta_{ij} =I(X_{ij}<C_{ij}$. $(X_{ij},\Delta_{ij})$ are the observed PFS for each patient.
    \item If $X_{ij}<t_{MRD}$ set $S_{ij}=1$ (MRD Positive).
    \item Return $(X_{ij},\Delta_{ij}, S_{ij})$
    
\end{enumerate} 

\subsection*{Simulation Verification}
For $S_{ij}$ defined in the previous section we have 
$$P(S_{ij}=0|Z_{ij}=z_{ij}) = p_0^{1-z_{ij}}p_1^{z_{ij}} $$
Let's define $b_z=b_0^{1-z_{ij}}b_1^{z_{ij}}$. Then we have
\begin{align*}
    \text{logit}(P(S_{ij}=0|Z_{ij}=z_{ij}))&= \text{log}\frac{P(S_{ij}=0|Z_{ij}=z_{ij})}{1-P(S_{ij}=0|Z_{ij}=z_{ij})}\\
    & = \text{log}\frac{p_0^{1-z_{ij}}p_1^{z_{ij}}}{1-p_0^{1-z_{ij}}p_1^{z_{ij}}}\\
    & = \text{log}\frac{\frac{b_z}{1+b_z}}{1-\frac{b_z}{b_z}}\\
    & = \text{log}\frac{b_z}{1+b_z}- \text{log}\frac{1}{1+b_z}\\
    & = \text{log}b_z\\
    & = \text{log}\left(b_0^{1-z_{ij}}b_1^{z_{ij}} \right)\\
    & = (1-z_{ij})\log b_0 z_{ij}\log b_1\\
    & = (1-z_{ij})\gamma_i +z_{ij}(\gamma_i+\alpha-i)\\
    & = \gamma_i + \alpha_iz_{ij}
\end{align*}

Thus, the marginal distribution of $S_{ij}|Z_{ij}$ is correctly specified in the simulation.

\newpage
\bibliography{refs.bib}

\end{document}